\documentstyle[12pt,epsfig,axodraw]{article}
\makeatletter
\def\vereq#1#2{\lower3pt\vbox{\baselineskip1.5pt \lineskip1.5pt
\ialign{$\m@th#1\hfill##\hfil$\crcr#2\crcr\sim\crcr}}}
\makeatother
\def\Ptar xvfR#1#2#3{Phys. Rev. {\bf #1} (#3) #2 }
\def\PRL#1#2#3{Phys. Rev. Lett. {\bf #1} (#3) #2 }
\def\PL#1#2#3{Phys. Lett. {\bf #1} (#3) #2 }

\def\NP#1#2#3{Nucl. Phys. {\bf #1} (#3) #2 }

\def\PTP#1#2#3{Prog. Theor. Phys. {\bf #1} (#3) #2 }

\begin{document}

\begin{titlepage}
\begin{flushright}
\end{flushright}

\vskip 1cm
\begin{center}
{\large\bf Unitarity in gauge symmetry breaking on orbifold}

\vskip 1cm {\normalsize Y.\ Abe, 
N.\ Haba, Y.\ Higashide, K.\ Kobayashi, M. Matsunaga}
\\
\vskip 0.5cm{\it Faculty of Engineering, Mie University, Tsu, Mie,
  514-8507, Japan}
\end{center}
\vskip .5cm
\vspace{.3cm}

\begin{abstract}

We study the unitarity bounds of the scattering amplitudes in 
 the 
 extra dimensional gauge theory 
 where the gauge symmetry is broken by the boundary 
 condition. 
The estimation of the amplitude of 
 the diagram including four massive gauge 
 bosons in the external lines 
 shows that the asymptotic power behavior of the 
 amplitude is canceled. 
The calculation will be done 
 in the 5 dimensional 
 standard model and the $SU(5)$ grand unified theory, 
 whose 5th dimensional coordinate 
 is compactified 
 on $S^1/Z_2$. 
The broken gauge theories through 
 the orbifolding preserve the 
 unitarity at high energies 
 similarly to 
 the broken gauge theories 
 where the gauge bosons obtain their 
 masses through the Higgs mechanism. 
We show that the contributions of 
 the Kaluza-Klein states
 play a crucial 
 role in conserving the unitarity. 

\end{abstract}
\end{titlepage}
\setcounter{footnote}{0}
\setcounter{page}{1}
\setcounter{section}{0}
\setcounter{subsection}{0}
\setcounter{subsubsection}{0}

\section{Introduction}

Much attention has been paid to gauge 
 theories in higher dimensions. 
In particular, theories 
 whose extra dimensional coordinates 
 are compactified on the orbifolds 
 have been studied 
 in the 
 standard model (SM) ({\it see, for examples,} \cite{CSM}) and 
 the grand unified 
 theories (GUTs) ({\it see, for 
 examples,} \cite{5d,5d1,5d2,flavor,Hall:2001zb}). 
One of the strongest motivations of 
 the higher dimensional gauge theory
 is the realization 
 of gauge symmetry reduction 
 by the boundary condition. 
The nontrivial parity 
 makes the gauge symmetries 
 be reduced ({\it see, for examples,} \cite{5d}). 
The broken gauge bosons 
 become the Kaluza-Klein (KK) states\cite{KK}, 
 which have masses of 
 $n/R$, where $R$ is the compactification 
 scale with the positive integer $n$. 
A solution of the triplet-doublet splitting 
 problem can be also realized 
 by this boundary condition\cite{5d}.

The mechanism of gauge symmetry breaking and 
 mass generation in the orbifold models 
 is very different from
 the Higgs mechanism. 
The nontrivial boundary condition set 
 by hand breaks the gauge symmetry from
 the beginning. 
One question, then, arises:
 Is the unitarity bounds on the asymptotic 
 high energy behavior of scattering 
 amplitude violated or not?\footnote{
There are some discussions 
 about unitarity
 in these theories in Ref.\cite{uni}.}

It is well-known that 
 the 4D gauge theories 
 with the explicit gauge 
 symmetry breakings 
 give rise to a power 
 behavior of the amplitude 
 and break unitarity 
 in general. 
The diagrams 
 containing massive gauge 
 bosons in the external lines 
 are the source of 
 the breaking of unitarity. 
While in the spontaneous broken 
 gauge theories where 
 the gauge bosons obtain
 their masses through
 the Higgs mechanism, 
 the power behaviors of the 
 amplitude are canceled
 by the 
 diagrams containing the Higgs 
 particles. 
The 3-point (Higgs-gauge-gauge) interactions 
 which is proportional 
 to the mass of the 
 gauge boson play a crucial 
 role in preserving the unitarity. 
The broken gauge theories 
 through the Higgs mechanism 
 can be applied above the 
 energy of the 
 compactification scale.

How is the situation in the extra 
 dimensional gauge theories 
 where the gauge symmetries 
 are broken by the boundary 
 conditions?  
The analysis of the unitarity bound 
 in the process of including two 
 massive gauge bosons in the external lines
 has been done in Ref.\cite{Hall:2001tn}.
They showed that the 
 power behavior 
 in the diagrams with two massive
 gauge bosons in the external
 lines is 
 canceled due to the 
 contributions of the KK states. 
We should check, however,  
 the diagrams with four massive 
 gauge bosons in the 
 external lines for 
 the accurate arguments 
 of the unitarity bound 
 in the broken gauge theory.  
The structure of the interactions among KK modes  
 is crucial for conserving the unitarity 
 in the diagrams with four massive 
 gauge bosons in the 
 external lines.

In this paper 
 we study the unitarity bound of the 
 extra dimensional gauge theory 
 where the gauge symmetry is broken by the boundary 
 condition. 
The gauge symmetries are broken 
 not by the Higgs mechanism but 
 by the boundary conditions. 
We analyze the unitarity bound in the process 
 of including four massive gauge bosons in the external lines. 
The result 
 shows the power behaviors of the 
 amplitude are canceled. 
The calculation will be done 
 in the 5 dimensional 
 SM and the $SU(5)$ GUT, 
 whose 5th dimensional coordinate 
 is compactified 
 on $S^1/Z_2$. 
The broken gauge theories 
 through 
 the orbifolding preserve the 
 unitarity at high energies similarly to 
 the 
 broken gauge theories 
 where the gauge bosons obtain their 
 masses through the Higgs mechanism. 
The structure of the interactions 
 among KK states
 play a crucial 
 role in conserving the unitarity.

%
%
%
\section{Gauge symmetry breaking by orbifolding}


At first, we show the setup. 
We consider the 5D gauge theory where 
 the gauge field exists 
 in the bulk. 
We denote the five dimensional coordinate 
 as $y$, which is 
 compactified on an $S^1/Z_2$ 
 orbifold. 
Under the $Z_2$ parity transformation, 
 $y \rightarrow -y$,  
 the gauge field $A_M(x^\mu,y)$ $(M= \mu(=0\sim3),5)$ 
 transforms as 
\begin{eqnarray}
A_\nu(x^\mu,y) &\to& A_\nu(x^\mu,-y) = PA_\nu(x^\mu,y)P^{-1},\\
A_5(x^\mu,y) &\to& A_5(x^\mu,-y) = -PA_5(x^\mu,y)P^{-1},
\end{eqnarray}
where $P$ is the operator of 
 $Z_2$ transformation. 
Two walls at $y=0$ and $\pi R$ are
 fixed points under $Z_2$ transformation. 
The physical space can be taken to be $0 \leq y \leq \pi R$. 
Here we take $Z_2$ as $P=1$, so that 
 the mode expansion of $A_\mu$ ($A_5$) for the 
 5D coordinate is given by series of cos- (sin-) functions. 
Besides, we consider the nontrivial 
 boundary conditions $T: y \rightarrow y+2\pi R$, 
 where the parity (reflection) 
 operator $P'$ around $y=\pi R$  
 is given by $P'=TP$. 
On this orbifold, the fields $A_\mu(x^\mu,y)$ and $A_5(x^\mu,y)$ 
 are divided into 
\begin{eqnarray}
\label{3}
A_{\nu +}(x^\mu,y) &=& \sum_{n=0}^{\infty} 
 \frac{1}{\sqrt{2^{\delta_{n,0}}\pi R}}\; A_{\nu +}^{(n)}
          (x^\mu)\; \cos \frac{ny}{R},\\
\label{4}
A_{\nu -}(x^\mu,y) &=& \sum_{n=0}^{\infty} 
  \frac{1}{\sqrt{\pi R}}\; A_{\nu -}^{(n+{1\over2})}
          (x^\mu)\; \cos \frac{(n+{1\over2})y}{R}, \\
A_{5 +}(x^\mu,y) &=& \sum_{n=0}^{\infty} 
  \frac{1}{\sqrt{\pi R}}\; A_{5 -}^{(n+{1\over2})}
          (x^\mu)\; \sin \frac{(n+{1\over2})y}{R}, \\
\label{2.6}
A_{5 -}(x^\mu,y) &=& \sum_{n=0}^{\infty} 
 \frac{1}{\sqrt{\pi R}}\; A_{5 +}^{(n+1)}
          (x^\mu)\; \sin \frac{(n+1)y}{R}, 
\end{eqnarray}
according to the eigenvalues, $\pm1$,  
 of the parity $P'$.

We can see that the gauge symmetry 
 is broken by taking the nontrivial 
 parity operator $P'$ in the gauge group basis. 
Considering $SU(3)_W \supset SU(2)_L\times U(1)_Y$ gauge theory, 
 the $Z_2$ parity operator, $P'= diag.(1,1,-1)$, 
 realizes the gauge reduction of 
 $SU(3)_W \rightarrow SU(2)_L\times U(1)_Y$\cite{Hall:2001zb}. 
It is because the gauge bosons of the cosset
 space $SU(3)_W/SU(2)_L\times U(1)_Y$ 
 become heavy through the mode expansions in Eqs.(\ref{3}) and (\ref{4}).
The way of generating gauge boson masses 
 is quite different from the
 Higgs mechanism.
We will examine the unitarity bound 
 in this theory in the next section. 
In the section 4, 
 we will consider the 5D $SU(5)$ theory with
 the $Z_2$ parity operator, $P'= diag.(-1,-1,-1,1,1)$, 
 which realizes the gauge reduction of 
 $SU(5) \rightarrow SU(3)_c \times SU(2)_L\times U(1)_Y$, 
 and estimate the unitarity bound of this theory.

\section{Unitarity in the SM on orbifold}

It is well-known that 
 massive gauge bosons in the external
 lines bring the unitarity bound to a crisis. 
At first, we show the notation 
 of massive vector boson in the external line. 
In the center-of-mass frame, 
 we take the initial momentum as 
 $p_1=(E,0,0,p)=E(1,0,0,\sqrt{1-{m^2\over E^2}})$ and
 $p_2=(E,0,0,-p)$, 
 and the final momentum as
 $k_1=(E,p\sin\theta,0,p\cos\theta)
     =E(1,\sqrt{1-{m^2\over E^2}}\sin\theta,0,
          \sqrt{1-{m^2\over E^2}}\cos\theta)$ and
 $k_2=(E,-p \sin \theta,0,-p \cos \theta)$,
 where $m$ is the gauge boson mass.
Then the longitudinal polarization vectors 
 become 
 $\epsilon_L(p_1)=({p\over m},0,0,{E \over m})
  ={E \over m}(\sqrt{1-{m^2\over E^2}},0,0,1)$ 
 and 
 $\epsilon_L(k_1)=({p\over m},{E \over m}\sin\theta,0,
   {E \over m}\cos\theta)
  ={E \over m}(\sqrt{1-{m^2\over E^2}},
    \sin\theta,0,\cos\theta)$.

Before examining 
 the unitarity bound in the orbifold model, 
 let us show briefly 
 the unitarity bound of 4D SM ({\it see, 
 for examples,} \cite{Cornwall}),
 where the $W$ and $Z$ gauge bosons obtain
 masses through the Higgs mechanism. 
Each tree diagram including 
 two massive gauge bosons in the
 external lines 
 is 
 of $O(E^2/m^2)$. 
It is because 
 the polarization vector $\epsilon^\mu(k)$
 is proportional to $\epsilon^\mu(k)\simeq k^\mu/m$
 in the high energy limit.  
However, in this order, 
 the unitarity 
 is maintained without Higgs 
 contributions. 
{}For examples, in the 
 process of  
 $e^+ e^- \rightarrow W^+ W^-$,   
 three diagrams $-$ 
 s-channel photon and $Z$ exchange,  
 t-channel neutrino exchange,  
 and four-point interaction processes $-$ 
 cancel the $O(E^2/m^2)$ behavior. 
The important point is that 
 this cancellation is realized 
 without the Higgs scalar contributions. 
Actually, 
 the cancellation of the power behavior 
 in the process including two massive
 gauge bosons in the external lines 
 can be 
 realized even when the gauge boson mass is not given by 
 the Higgs 
 mechanism. 

Then how about 
 the amplitude including 
 four massive gauge bosons in the
 external line?
This case seems to induce power behavior of $O(E^4/m^4)$, 
 naively. 
{}For examples, 
 let us see 
 the process of 
 $W^+ W^- \rightarrow W^+ W^-$. 
There are following three diagrams 
 for this process (Figs.1(a)$\sim$(c)). 
\begin{figure}[hbt]
  \begin{picture}(410,130)(15,0)
  \setlength{\unitlength}{1pt}
%
%
    \Photon(60,35)(95,55){4}{5}
    \Photon(60,115)(95,95){4}{5}
    \Photon(95,95)(95,55){4}{5}
    \Vertex(95,55){1.5}
    \Vertex(95,95){1.5}
    \Photon(95,95)(130,115){4}{5}
    \Photon(95,55)(130,35){4}{5}
    \put(50,20){$W^+$}
    \put(130,20){$W^-$}
    \put(50,120){$W^+$}
    \put(130,120){$W^-$}
    \put(105,75){$\gamma, Z$}
    \put(75,5){Fig.1(a)}
%
%
    \Photon(190,45)(215,75){4}{5}
    \Photon(190,105)(215,75){4}{5}
    \Photon(215,75)(255,75){4}{5}
    \Vertex(215,75){1.5}
    \Vertex(255,75){1.5}
    \Photon(280,45)(255,75){4}{5}
    \Photon(280,105)(255,75){4}{5}
    \put(190,30){$W^+$}
    \put(270,30){$W^-$}
    \put(190,110){$W^+$}
    \put(270,110){$W^-$}
    \put(225,60){$\gamma, Z$}
    \put(215,5){Fig.1(b)}
%
%
    \Photon(325,45)(360,75){4}{5}
    \Photon(325,105)(360,75){4}{5}
    \Photon(395,45)(360,75){4}{5}
    \Photon(395,105)(360,75){4}{5}
    \Vertex(360,75){1.5}
    \put(325,30){$W^+$}
    \put(385,30){$W^-$}
    \put(325,110){$W^+$}
    \put(385,110){$W^-$}
    \put(345,5){Fig.1(c)}
  \end{picture}
\end{figure}
They cancel each other up to the order 
 of $O(E^4/m^4)$. 
However, there still remains
 the divergence of $O(E^2/m^2)$. 
This brings about the violation
 of the unitarity, and 
 the theory including massive gauge
 bosons must be the effective
 theory bellow the energy scale of $m$. 
However, it is well-known that 
 the SM, where the gauge bosons 
 obtain masses through the Higgs mechanism, 
 has additional following 
 two diagrams (Figs1(d),(e)). 
\begin{figure}[hbt]
  \begin{picture}(410,130)(0,0)
  \setlength{\unitlength}{1pt}
%
%
    \Photon(60,35)(95,55){4}{5}
    \Photon(60,115)(95,95){4}{5}
    \DashArrowLine(95,55)(95,95){3}
    \Vertex(95,55){1.5}
    \Vertex(95,95){1.5}
    \Photon(95,95)(130,115){4}{5}
    \Photon(95,55)(130,35){4}{5}
    \put(50,20){$W^+$}
    \put(130,20){$W^-$}
    \put(50,120){$W^+$}
    \put(130,120){$W^-$}
    \put(105,75){$h^0$}
    \put(75,5){Fig.1(d)}
%
%
    \Photon(230,45)(255,75){4}{5}
    \Photon(230,105)(255,75){4}{5}
    \DashArrowLine(255,75)(295,75){3}
    \Vertex(255,75){1.5}
    \Vertex(295,75){1.5}
    \Photon(295,75)(320,105){4}{5}
    \Photon(295,75)(320,45){4}{5}
    \put(230,30){$W^+$}
    \put(310,30){$W^-$}
    \put(230,110){$W^+$}
    \put(310,110){$W^-$}
    \put(275,60){$h^0$}
    \put(255,5){Fig.1(e)}
  \end{picture}
\end{figure}
Then as is seen in 
 the Table.I, $O(E^2/m^2)$ becomes
 vanished by adding all 
 diagrams (Figs.1(a)$\sim$(e)). 
This result is expected from the so-called 
 equivalence theorem, 
 which means that 
 the amplitude of the diagram with the massive gauge boson 
 in the external lines is, up to $O(m/E)$ corrections, 
 equal to that of 
 diagram of the scalar (would-be NG boson) in the external
 lines instead of massive gauge boson. 
Thus, the spontaneous broken 
 gauge theories are expected to be 
 always protected from 
 the unitarity violation. 
They can be the true theory 
 above the energy scale of gauge boson mass. 
On the other hand, 
 it is not the case for
 the theory with explicit symmetry breaking. 
If the gauge boson mass is not coming from
 the Higgs mechanism, the unitarity 
 is not necessarily guaranteed. 
It is because 
 the 3-point vertex, $(W-W-h^0)$, with the coupling 
 proportional to the gauge boson mass
 plays a crucial role in the 
 above cancellation.
So if there are no Higgs scalars, 
 there do not exist the diagrams
 of Figs.1(d) and (e), where the 
 Higgs mechanism does not work and the absence of 
 3-point vertex proportional to 
 the gauge boson mass induces the 
 power behavior in general. 
%
%
%
\begin{table}[h]
\caption[table-1]{The coefficients of the 
amplitude of $W^+W^-\rightarrow W^+W^-$ in
 Figs.1(a)$\sim$(e) in the 4D SM. 
Both $O(E^4/m^4)$ and $O(E^2/m^2)$ 
 are canceled among Figs.1(a)$\sim$(e).}
\begin{center}
\begin{tabular}{|c||c|c|}     \hline
\multicolumn{1}{|c|} {}& {${(ig^2E^4 / m^4)\times}$} 
        & {${(ig^2E^2 / m^2)\times}$}  \\ \hline
\multicolumn{1}{|c|} {Fig.1(a)} & {$-4\cos \theta$} &
{$-\cos \theta$} \\ \hline
\multicolumn{1}{|c|} {Fig.1(b)} & {$3-2\cos \theta -\cos^2\theta$} &
{$-3/2+(15/2)\cos \theta$} \\ \hline
\multicolumn{1}{|c|} {Fig.1(c)} & {$-3+6\cos \theta +\cos^2\theta$} &
{$2-6\cos \theta$} \\ \hline
\multicolumn{1}{|c|} {Figs.1(d),(e)} & {$-$} &
{$-1/2-(1/2)\cos \theta$} \\ \hline
\end{tabular}
\end{center}
\label{table-1}
\end{table}

Now let us consider the unitarity bounds in 
 the 5D SM on the orbifold, 
 where the gauge bosons are obtaining 
 KK masses through
 the orbifolding. 
In this case, we should take into account 
 the contribution of virtual photon and 
 $Z$ boson of the KK excited states. 
It should be noticed that 
 gauge couplings 
 among KK states 
 obey some kind of selection rule. 
We take the ``gauge'' where 
 the 5th gauge field  
 $A_5 (x^\mu, y)$ is 
 {\it gauged away}\footnote{This ``gauge fixing'' 
 will be discussed 
 in the section 5.}.  
{}For example, 
 let us consider the process of 
 of $W^{(1/2)+}W^{(1/2)-}\rightarrow W^{(1/2)+}W^{-(1/2)-}$. 
We can easily see that 
 zero mode and first excited photon and $Z$ 
 can only have the 
 3-point coupling (pair creation) of 
 $W^{(1/2)+}W^{(1/2)-}$. 
Then there are following three diagrams 
 in the 5D SM: 
\begin{figure}[hbt]
  \begin{picture}(410,130)(20,0)
  \setlength{\unitlength}{1pt}
%
%
    \Photon(60,35)(95,55){4}{5}
    \Photon(60,115)(95,95){4}{5}
    \Photon(95,95)(95,55){4}{5}
    \Vertex(95,55){1.5}
    \Vertex(95,95){1.5}
    \Photon(95,95)(130,115){4}{5}
    \Photon(95,55)(130,35){4}{5}
    \put(50,20){$W^{(1/2)+}$}
    \put(130,20){$W^{(1/2)-}$}
    \put(50,120){$W^{(1/2)+}$}
    \put(130,120){$W^{(1/2)-}$}
    \put(45,75){$\gamma^{(0)}, Z^{(0)}$}
    \put(105,75){$\gamma^{(1)}, Z^{(1)}$}
    \put(75,5){Fig.2(a)}
%
%
    \Photon(190,45)(210,75){4}{5}
    \Photon(190,105)(210,75){4}{5}
    \Photon(210,75)(260,75){4}{5}
    \Vertex(210,75){1.5}
    \Vertex(260,75){1.5}
    \Photon(280,45)(260,75){4}{5}
    \Photon(280,105)(260,75){4}{5}
    \put(190,30){$W^{(1/2)+}$}
    \put(270,30){$W^{(1/2)-}$}
    \put(190,110){$W^{(1/2)+}$}
    \put(270,110){$W^{(1/2)-}$}
    \put(215,60){$\gamma^{(0)}, Z^{(0)}$}
    \put(215,85){$\gamma^{(1)}, Z^{(1)}$}
    \put(215,5){Fig.2(b)}
%
%
    \Photon(325,45)(360,75){4}{5}
    \Photon(325,105)(360,75){4}{5}
    \Photon(395,45)(360,75){4}{5}
    \Photon(395,105)(360,75){4}{5}
    \Vertex(360,75){1.5}
    \put(325,30){$W^{(1/2)+}$}
    \put(385,30){$W^{(1/2)-}$}
    \put(325,110){$W^{(1/2)+}$}
    \put(385,110){$W^{(1/2)-}$}
    \put(345,5){Fig.2(c)}
  \end{picture}
\end{figure}

We remind that   
 the 5D gauge coupling, $g_5$, 
 has mass dimension $-1/2$, 
 which is related to the 4D gauge coupling, $g_4$, 
 as $g_4=g_5/\sqrt{2\pi R}$.  
Then 
 the couplings of $(\gamma^{(0)}-W^{(1/2)+}-W^{(1/2)-})$ and
 $(Z^{(0)}-W^{(1/2)+}-W^{(1/2)-})$ are 
\begin{eqnarray}
\label{g51}
g_5 \int_{0}^{2\pi R} dy 
 \left({1\over \sqrt{\pi R}}\cos{y \over 2R} \right)^2 {1\over
 \sqrt{2\pi R}}=g_4,
\end{eqnarray}
while 
 the couplings of $(\gamma^{(1)}-W^{(1/2)+}-W^{(1/2)-})$ and
 $(Z^{(1)}-W^{(1/2)+}-W^{(1/2)-})$ are 
\begin{eqnarray}
\label{g52}
g_5 \int_{0}^{2\pi R} dy 
 \left({1\over \sqrt{\pi R}}\cos{y \over 2R} \right)^2 
 \left({1\over \sqrt{\pi R}}\cos{y \over R} \right) 
 ={g_4 \over \sqrt{2}}.
\end{eqnarray}
Thus, 
 the amplitude of Fig.2(a) (Fig.2(b)) becomes 
 3/2 times that of Fig.1(a) (Fig.1(b)), 
 by including both contributions of $\gamma^{(0)}, Z^{(0)}$ 
 and $\gamma^{(1)}, Z^{(1)}$. 
As for the four point vertex in Fig.2(c),
 the coupling is 
 given by 
\begin{eqnarray}
\label{g53}
i g_5^2 \int_{0}^{2\pi R} dy 
 \left({1\over \sqrt{\pi R}}\cos{y \over 2R} \right)^4 
 ={3\over2}i g_4^2. 
\end{eqnarray}
This means that 
 the amplitude of Fig.2(c) becomes 
 3/2 times that of Fig.1(c). 
The results is shown in Table II. 
%
%
%
\begin{table}[h]
\caption[table-2]{The coefficients of the 
amplitude of $W^{(1/2)+}W^{(1/2)-}\rightarrow W^{(1/2)+}W^{(1/2)-}$ in
 Figs.2(a)$\sim$(c) in the 5D SM. 
Power behaviors of $O(E^4/m^4)$ and $O(E^2/m^2)$ 
 are canceled among Figs.2(a)$\sim$(c).}
\begin{center}
\begin{tabular}{|c||c|c|}     \hline
\multicolumn{1}{|c|} {}& {${(ig^2E^4 / m^4)\times}$} 
        & {${(ig^2E^2 / m^2)\times}$}  \\ \hline
\multicolumn{1}{|c|} {Fig.2(a)} & {$-6\cos \theta$} &
{$-2\cos \theta$} \\ \hline
\multicolumn{1}{|c|} {Fig.2(b)} & {$9/2-3\cos \theta -(3/2)\cos^2\theta$}
 &
{$-3+11\cos \theta$} \\ \hline
\multicolumn{1}{|c|} {Fig.2(c)} & {$-9/2+9\cos \theta 
+(3/2)\cos^2\theta$} &
{$3-9\cos \theta$} \\ \hline
\end{tabular}
\end{center}
\label{table-2}
\end{table}
Both of the power behaviors of 
 $O(E^4/m^4)$ and $O(E^2/m^2)$ are canceled 
 although there are no Higgs contribution 
 as the 4D SM (Figs.1(d),(e)). 
Table II suggests that 
 the KK-modes play an important 
 role for preserving the unitarity. 
They realize the cancellation of 
 the power behavior of $O(E^2/m^2)$ 
 as the Higgs scalars do 
 in the spontaneous
 breaking gauge theories as well as 
 that of $O(E^4/m^4)$. 

\section{Unitarity in the GUT on orbifold}

Next, 
 we consider the 5D $SU(5)$ theory with
 the $Z_2$ parity operator, $P'= diag.(-1,-1,-1,1,1)$, 
 which realizes the gauge reduction of 
 $SU(5) \rightarrow SU(3)_c \times SU(2)_L\times U(1)_Y$ \cite{5d}. 
The analysis of the unitarity bound 
 in the process of including two 
 massive gauge bosons in the external lines
 has been done in Ref.\cite{Hall:2001tn}.
They estimate the process of 
 $H_D H_D^* \rightarrow X X^*$. 
It has been known that 
 three diagrams $-$ 
 s-channel $U(1)_Y$ and $SU(2)_L$ gauge boson exchange,    
 t-channel colored Higgs exchange, and 
 four-point interaction processes $-$
 cancel the 
$O(E^2/m^2)$ behavior in the usual 
 4D $SU(5)$ GUT. 
On the other hand, 
 in the corresponding process in 5D $SU(5)$
 GUT, $H_D H_D^* \rightarrow X^{(1/2)} X^{(1/2)*}$, 
 the $O(E^2/m^2)$ divergence is canceled 
 among the s-channel exchange diagrams of zero mode and 
 KK mode. 
The Higgs boson $H_D$ is the brane-localized field on $y=\pi R$.

How about 
 the unitarity bound in the process 
 of including four massive gauge bosons 
 in the external lines? 
As is shown in the previous 
 section, this process is 
 more important for 
 the precise arguments of 
 the unitarity bounds. 
Let us consider the processes, 
 $X X^{*} \rightarrow X X^{*}$ in the 4D GUT and 
 $X^{(1/2)} X^{(1/2)*} \rightarrow X^{(1/2)} X^{(1/2)*}$
 of the 5D GUT. 
This case seems to induce power behavior of $O(E^4/m^4)$ 
 naively as described in the previous section. 
In the 4D $SU(5)$ GUT, 
 there are following three diagrams 
 of this process (Figs.3(a)$\sim$(c)). 
\begin{figure}[h]
  \begin{picture}(410,130)(15,0)
  \setlength{\unitlength}{1pt}
%
%
    \Photon(60,35)(95,55){4}{5}
    \Photon(60,115)(95,95){4}{5}
    \Photon(95,95)(95,55){4}{5}
    \Vertex(95,55){1.5}
    \Vertex(95,95){1.5}
    \Photon(95,95)(130,115){4}{5}
    \Photon(95,55)(130,35){4}{5}
    \put(50,20){$X$}
    \put(130,20){$X^*$}
    \put(50,120){$X$}
    \put(130,120){$X^*$}
    \put(55,75){$A_3,A_8$}
    \put(105,75){$W_3,B$}
    \put(75,5){Fig.3(a)}
%
%
    \Photon(190,45)(215,75){4}{5}
    \Photon(190,105)(215,75){4}{5}
    \Photon(215,75)(255,75){4}{5}
    \Vertex(215,75){1.5}
    \Vertex(255,75){1.5}
    \Photon(280,45)(255,75){4}{5}
    \Photon(280,105)(255,75){4}{5}
    \put(190,30){$X$}
    \put(270,30){$X^*$}
    \put(190,110){$X$}
    \put(270,110){$X^*$}
    \put(220,60){$A_3,A_8$}
    \put(220,85){$W_3,B$}
    \put(215,5){Fig.3(b)}
%
%
    \Photon(325,45)(360,75){4}{5}
    \Photon(325,105)(360,75){4}{5}
    \Photon(395,45)(360,75){4}{5}
    \Photon(395,105)(360,75){4}{5}
    \Vertex(360,75){1.5}
    \put(325,30){$X$}
    \put(385,30){$X^*$}
    \put(325,110){$X$}
    \put(385,110){$X^*$}
    \put(345,5){Fig.3(c)}
  \end{picture}
\end{figure}
Where the 
 propagators, $A_3$, $A_8$, $W_3$, and $B$, 
 stand for the diagonal 
 elements of the gauge fields of the $SU(5)$, 
 corresponding to $SU(3)_c$, $SU(3)_c$, $SU(2)_L$, 
 and $U(1)_Y$ components, respectively. 
Figures 3(a)$\sim$(c) 
 cancel each other up to the order 
 of $O(E^4/m^4)$. 
However, there still remains
 the divergence of $O(E^2/m^2)$. 
This divergence is 
 canceled by additional two 
 diagrams including 
 adjoint Higgs $(\Sigma)$ 
 contributions. 
\begin{figure}[h]
  \begin{picture}(410,130)(0,0)
  \setlength{\unitlength}{1pt}
%
%
    \Photon(60,35)(95,55){4}{5}
    \Photon(60,115)(95,95){4}{5}
    \DashArrowLine(95,55)(95,95){3}
    \Vertex(95,55){1.5}
    \Vertex(95,95){1.5}
    \Photon(95,95)(130,115){4}{5}
    \Photon(95,55)(130,35){4}{5}
     \put(50,20){$X$}
    \put(130,20){$X^*$}
    \put(50,120){$X$}
    \put(130,120){$X^*$}
    \put(55,75){$\Sigma_3,\Sigma_8$}
    \put(105,75){$\Sigma_{W_3},\Sigma_B$}
    \put(75,5){Fig.3(d)}
%
%
    \Photon(230,45)(255,75){4}{5}
    \Photon(230,105)(255,75){4}{5}
    \DashArrowLine(255,75)(295,75){3}
    \Vertex(255,75){1.5}
    \Vertex(295,75){1.5}
    \Photon(295,75)(320,105){4}{5}
    \Photon(295,75)(320,45){4}{5}
    \put(230,30){$X$}
    \put(310,30){$X^*$}
    \put(230,110){$X$}
    \put(310,110){$X^*$}
    \put(260,60){$\Sigma_3,\Sigma_8$}
    \put(255,85){$\Sigma_{W_3},\Sigma_B$}
    \put(255,5){Fig.3(e)}
  \end{picture}
\end{figure}
Where 
 the propagators, $\Sigma_3$, $\Sigma_8$, $\Sigma_{W_3}$, and
 $\Sigma_B$,
 stand for the diagonal 
 elements of the adjoint Higgs fields of the $SU(5)$.

As shown in 
 the Table III, $O(E^2/m^2)$ becomes
 vanished by adding all 
 diagrams (Figs.3(a)$\sim$(e)). 
The 3-point vertex, $(X-X-\Sigma_0)$, with the coupling 
 proportional to the gauge boson mass 
 plays a crucial role in the 
  cancellation.
%
%
%
\begin{table}[h]
\caption[table-3]{The coefficients of the 
amplitude of $XX^*\rightarrow XX^*$ in
 Figs.3(a)$\sim$(e) in the 4D $SU(5)$ GUT. 
Both $O(E^4/m^4)$ and $O(E^2/m^2)$ 
 are canceled among Figs.3(a)$\sim$(e).}
\begin{center}
\begin{tabular}{|c||c|c|}     \hline
\multicolumn{1}{|c|} {}& {${(ig^2E^4 / m^4)\times}$} 
        & {${(ig^2E^2 / m^2)\times}$}  \\ \hline
\multicolumn{1}{|c|} {Fig.3(a)} & {$-4\cos \theta$} &
{$-$} \\ \hline
\multicolumn{1}{|c|} {Fig.3(b)} & {$3-2\cos \theta -\cos^2\theta$} &
{$8\cos \theta$} \\ \hline
\multicolumn{1}{|c|} {Fig.3(c)} & {$-3+6\cos \theta +\cos^2\theta$} &
{$2-6\cos \theta$} \\ \hline
\multicolumn{1}{|c|} {Figs.3(d),(e)} & {$-$} &
{$-2-2\cos \theta$} \\ \hline
\end{tabular}
\end{center}
\label{table-3}
\end{table}

Then how about 
 the 5D $SU(5)$ on the orbifold?
Let us consider the process, 
 $X^{(1/2)} X^{(1/2)*} \rightarrow X^{(1/2)} X^{(1/2)*}$
 of the 5D GUT. 
In this case, we should take into account 
 not only the contributions of zero modes (Figs.4(a) and (b)) 
 but also 
 the KK excited states (Figs.4(d) and (e)). 
%
%
\begin{figure}[hbt]
  \begin{picture}(410,130)(20,0)
  \setlength{\unitlength}{1pt}
%
%
    \Photon(60,35)(95,55){4}{5}
    \Photon(60,115)(95,95){4}{5}
    \Photon(95,95)(95,55){4}{5}
    \Vertex(95,55){1.5}
    \Vertex(95,95){1.5}
    \Photon(95,95)(130,115){4}{5}
    \Photon(95,55)(130,35){4}{5}
    \put(50,20){$X^{(1/2)}$}
    \put(130,20){$X^{(1/2)*}$}
    \put(50,120){$X^{(1/2)}$}
    \put(130,120){$X^{(1/2)*}$}
    \put(45,75){$A_3^{(0)},A_8^{(0)}$}
    \put(105,75){$W_3^{(0)},B^{(0)}$}
    \put(75,5){Fig.4(a)}
%
%
    \Photon(190,45)(210,75){4}{5}
    \Photon(190,105)(210,75){4}{5}
    \Photon(210,75)(260,75){4}{5}
    \Vertex(210,75){1.5}
    \Vertex(260,75){1.5}
    \Photon(280,45)(260,75){4}{5}
    \Photon(280,105)(260,75){4}{5}
    \put(190,30){$X^{(1/2)}$}
    \put(270,30){$X^{(1/2)*}$}
    \put(190,110){$X^{(1/2)}$}
    \put(270,110){$X^{(1/2)*}$}
    \put(212,58){$A_3^{(0)},A_8^{(0)}$}
    \put(215,85){$W_3^{(0)},B^{(0)}$}
    \put(215,5){Fig.4(b)}
%
%
    \Photon(325,45)(360,75){4}{5}
    \Photon(325,105)(360,75){4}{5}
    \Photon(395,45)(360,75){4}{5}
    \Photon(395,105)(360,75){4}{5}
    \Vertex(360,75){1.5}
    \put(325,30){$X^{(1/2)}$}
    \put(385,30){$X^{(1/2)*}$}
    \put(325,110){$X^{(1/2)}$}
    \put(385,110){$X^{(1/2)*}$}
    \put(345,5){Fig.4(c)}
  \end{picture}
\end{figure}
%
%
%
%
%
\begin{figure}[hbt]
  \begin{picture}(410,130)(0,0)
  \setlength{\unitlength}{1pt}
    \Photon(60,35)(95,55){4}{5}
    \Photon(60,115)(95,95){4}{5}
    \Photon(95,95)(95,55){4}{5}
    \Vertex(95,55){1.5}
    \Vertex(95,95){1.5}
    \Photon(95,95)(130,115){4}{5}
    \Photon(95,55)(130,35){4}{5}
    \put(50,20){$X^{(1/2)}$}
    \put(130,20){$X^{(1/2)*}$}
    \put(50,120){$X^{(1/2)}$}
    \put(130,120){$X^{(1/2)*}$}
    \put(45,75){$A_3^{(1)},A_8^{(1)}$}
    \put(105,75){$W_3^{(1)},B^{(1)}$}
    \put(75,5){Fig.4(d)}
%
%
    \Photon(230,45)(250,75){4}{5}
    \Photon(230,105)(250,75){4}{5}
    \Photon(250,75)(300,75){4}{5}
    \Vertex(250,75){1.5}
    \Vertex(300,75){1.5}
    \Photon(300,75)(320,105){4}{5}
    \Photon(300,75)(320,45){4}{5}
    \put(230,30){$X^{(1/2)}$}
    \put(310,30){$X^{(1/2)*}$}
    \put(230,110){$X^{(1/2)}$}
    \put(310,110){$X^{(1/2)*}$}
    \put(255,58){$A_3^{(1)},A_8^{(1)}$}
    \put(255,85){${W_3}^{(1)},B^{(1)}$}
    \put(255,5){Fig.4(e)}
  \end{picture}
\end{figure}
By Eqs.(\ref{g51})$\sim$(\ref{g53}), 
 the amplitudes of Fig.4(a)$\sim$(e) become 
 3/2 times that of Fig.3(a)$\sim$(c). 
Then the amplitudes from Figs.4(a)$\sim$(c) 
 are shown in Table IV.  
%
%
%
\begin{table}[h]
\caption[table-4]{The coefficients of the 
 amplitude of $X^{(1/2)}X^{(1/2)*}\rightarrow X^{(1/2)}X^{(1/2)*}$ in
 Figs.4(a)$\sim$(e) in the 5D $SU(5)$ GUT. 
Power behaviors $O(E^4/m^4)$ and $O(E^2/m^2)$ 
 are canceled among Figs.4(a)$\sim$(e).}
\begin{center}
\begin{tabular}{|c||c|c|}     \hline
\multicolumn{1}{|c|} {}& {${(ig^2E^4 / m^4)\times}$} 
        & {${(ig^2E^2 / m^2)\times}$}  \\ \hline
\multicolumn{1}{|c|} {Figs.4(a),(d)} & {$-6\cos \theta$} &
{$-2\cos \theta$} \\ \hline
\multicolumn{1}{|c|} {Figs.4(b),(e)} & 
{$9/2-3\cos \theta -(3/2)\cos^2\theta$}
 &
{$-3+11\cos \theta$} \\ \hline
\multicolumn{1}{|c|} {Fig.4(c)} & {$-9/2+9\cos \theta 
+(3/2)\cos^2\theta$} &
{$3-9\cos \theta$} \\ \hline
\end{tabular}
\end{center}
\label{table-4}
\end{table}
As in the 4D GUT, both 
 $O(E^4/m^4)$ and 
 $O(E^2/m^2)$ are 
 canceled. 
In more general processes, 
 $X^{(n/2)}X^{(n/2)*}\rightarrow
 X^{(n/2)}X^{(n/2)*}$ $(n=1,2,..)$, 
 the power behaviors of 
 $O(E^4/m^4)$ and 
 $O(E^2/m^2)$ are 
 also canceled. 
Where 
 the propagators in 
 Figs.4(d) and (e) are replaced 
 as  $A_3^{(1)},A_8^{(1)},W_3^{(1)},B^{(1)}
 \rightarrow A_3^{(n)},A_8^{(n)},W_3^{(n)},B^{(n)}$, 
 and the amplitudes are exactly same as 
 those of the Table IV.

How about other 
 processes where masses of the final gauge 
 bosons are different from those of 
 the initial ones? 
Let us estimate 
 the amplitude, $X^{(1/2)}X^{(1/2)*}\rightarrow
 X^{(3/2)}X^{(3/2)*}$, for examples. 
There are four diagrams as in Figs.5(a)$\sim$(d). 
\begin{figure}[hbt]
  \begin{picture}(410,130)(20,0)
  \setlength{\unitlength}{1pt}
%
%
    \Photon(60,35)(95,55){4}{5}
    \Photon(60,115)(95,95){4}{5}
    \Photon(95,95)(95,55){4}{5}
    \Vertex(95,55){1.5}
    \Vertex(95,95){1.5}
    \Photon(95,95)(130,115){4}{5}
    \Photon(95,55)(130,35){4}{5}
    \put(50,20){$X^{(1/2)}$}
    \put(130,20){$X^{(1/2)*}$}
    \put(50,120){$X^{(3/2)}$}
    \put(130,120){$X^{(3/2)*}$}
    \put(45,75){$A_3^{(0)},A_8^{(0)}$}
    \put(105,75){$W_3^{(0)},B^{(0)}$}
    \put(75,5){Fig.5(a)}
%
%
    \Photon(190,45)(210,75){4}{5}
    \Photon(190,105)(210,75){4}{5}
    \Photon(210,75)(260,75){4}{5}
    \Vertex(210,75){1.5}
    \Vertex(260,75){1.5}
    \Photon(280,45)(260,75){4}{5}
    \Photon(280,105)(260,75){4}{5}
    \put(190,30){$X^{(1/2)}$}
    \put(270,30){$X^{(1/2)*}$}
    \put(190,110){$X^{(3/2)}$}
    \put(270,110){$X^{(3/2)*}$}
    \put(212,58){$A_3^{(1)},A_8^{(1)}$}
    \put(215,85){$W_3^{(1)},B^{(1)}$}
    \put(215,5){Fig.5(b)}
%
%
    \Photon(325,45)(360,75){4}{5}
    \Photon(325,105)(360,75){4}{5}
    \Photon(395,45)(360,75){4}{5}
    \Photon(395,105)(360,75){4}{5}
    \Vertex(360,75){1.5}
    \put(325,30){$X^{(1/2)}$}
    \put(385,30){$X^{(1/2)*}$}
    \put(325,110){$X^{(3/2)}$}
    \put(385,110){$X^{(3/2)*}$}
    \put(345,5){Fig.5(c)}
  \end{picture}
\end{figure}
\begin{figure}[hbt]
  \begin{picture}(350,120)(70,0)
  \setlength{\unitlength}{1pt}
%
%
    \Photon(230,45)(250,75){4}{5}
    \Photon(230,105)(250,75){4}{5}
    \Photon(250,75)(300,75){4}{5}
    \Vertex(250,75){1.5}
    \Vertex(300,75){1.5}
    \Photon(300,75)(320,105){4}{5}
    \Photon(300,75)(320,45){4}{5}
    \put(230,30){$X^{(1/2)}$}
    \put(310,30){$X^{(1/2)*}$}
    \put(230,110){$X^{(3/2)}$}
    \put(310,110){$X^{(3/2)*}$}
    \put(255,58){$A_3^{(2)},A_8^{(2)}$}
    \put(255,85){$W_3^{(2)},B^{(2)}$}
    \put(255,5){Fig.5(d)}
  \end{picture}
\end{figure}
By Eq.(\ref{g51}), 
 and using 
\begin{eqnarray}
&& g_5 \int_0^{2\pi R} dy 
 \left({1\over \sqrt{\pi R}}\cos{3y \over 2R} \right)^2 
{1\over \sqrt{2\pi R}}=g_4, \\
&& g_5 \int_0^{2 \pi R} dy
 \left({1\over \sqrt{\pi R}}\cos{y \over 2R} \right)
 \left({1\over \sqrt{\pi R}}\cos{3y \over 2R} \right)
 \left({1\over \sqrt{\pi R}}\cos{y \over R} \right) ={g_4 \over \sqrt{2}}, \\
&& g_5 \int_0^{2 \pi R} dy
 \left({1\over \sqrt{\pi R}}\cos{y \over 2R} \right)
 \left({1\over \sqrt{\pi R}}\cos{3y \over 2R} \right)
 \left({1\over \sqrt{\pi R}}\cos{2y \over R} \right) ={g_4 \over \sqrt{2}}, \\
&&i g_5^2 \int_{0}^{2\pi R} dy 
 \left({1\over \sqrt{\pi R}}\cos{y \over 2R} \right)^2 
 \left({1\over \sqrt{\pi R}}\cos{3y \over 2R} \right)^2 
 =i g_4^2 ,
\end{eqnarray}
the amplitudes of Fig.5(a)$\sim$(d) are 
 calculated as Table V. 
%
%
%
\begin{table}[h]
\caption[table-5]{The coefficients of the 
 amplitude of $X^{(1/2)}X^{(1/2)*}\rightarrow X^{(3/2)}X^{(3/2)*}$ in
 Figs.5(a)$\sim$(d) in the 5D $SU(5)$ GUT. 
Power behaviors $O(E^4/m^4)$ and $O(E^2/m^2)$ 
 are canceled among Figs.5(a)$\sim$(d).}
\begin{center}
\begin{tabular}{|c||c|c|}     \hline
\multicolumn{1}{|c|} {}& {${(ig^2E^4 / m^4)\times}$} 
        & {${(ig^2E^2 / m^2)\times}$}  \\ \hline
\multicolumn{1}{|c|} {Fig.5(a)} & {$-4\cos \theta$} &
{$-$} \\ \hline
\multicolumn{1}{|c|} {Figs.5(b),(d)} & 
{$3-2\cos \theta -\cos^2\theta$}
 &
{$-10+30\cos \theta$} \\ \hline
\multicolumn{1}{|c|} {Fig.5(c)} & {$-3+6\cos \theta 
+\cos^2\theta$} &
{$10-30\cos \theta$} \\ \hline
\end{tabular}
\end{center}
\label{table-5}
\end{table}
This case also suggests 
 the cancellation of 
 the power behaviors of $O(E^4/m^4)$ and 
 $O(E^2/m^2)$. 
Furthermore, the amplitudes of 
 the processes, 
 $X^{(n/2)}X^{(n/2)*}\rightarrow
 X^{(l/2)}X^{(l/2)*}$ $(n,l=1,2,..$ with $n \neq l)$, 
 are shown in 
 Table VI. 
Where 
 the propagators in 
 Figs.5(b),(d) are replaced 
 as  $A_3^{(1,2)},A_8^{(1,2)},W_3^{(1,2)},B^{(1,2)}
 \rightarrow A_3^{((n\pm l)/2)},A_8^{((n\pm l)/2)},
 W_3^{((n\pm l)/2)},B^{((n\pm l)/2)}$. 
In these cases the power behaviors of 
 $O(E^4/m^4)$ and 
 $O(E^2/m^2)$ are 
 also canceled. 
\begin{table}[h]
\caption[table-6]{The coefficients of the 
 amplitude of $X^{(n/2)}X^{(n/2)*}\rightarrow X^{(l/2)}X^{(l/2)*}$ 
 $(n,l=1,2,..$ with $n \neq l)$, in
 the 5D $SU(5)$ GUT. 
Power behaviors $O(E^4/m^4)$ and $O(E^2/m^2)$ 
 are canceled. }
\begin{center}
\begin{tabular}{|c||c|c|}     \hline
\multicolumn{1}{|c|} {}& {${(ig^2E^4 / m^4)\times}$} 
        & {${(ig^2E^2 / m^2)\times}$}  \\ \hline
\multicolumn{1}{|c|} {s-channel} & {$-4\cos \theta$} &
{$-$} \\ \hline
\multicolumn{1}{|c|} {t-channel} & 
{$3-2\cos \theta -\cos^2\theta$}
 &
{$\left(1+{l^2 \over n^2}\right)(-1+3\cos \theta )$} \\ \hline
\multicolumn{1}{|c|} {4-point} & 
{$-3+6\cos \theta +\cos^2\theta$} &
{$\left(1+{l^2 \over n^2}\right)(1-3\cos \theta )$} \\ \hline
\end{tabular}
\end{center}
\label{table-6}
\end{table}

Above calculations suggest 
 that 
 the 5D GUT theory with the 
 gauge symmetry violation 
 through the boundary condition
 also preserves 
 the unitarity.

\section{Summary and discussion}

We have studied the unitarity bounds of the 
 extra dimensional gauge theory 
 where the gauge symmetry is broken by the boundary 
 condition. 
The gauge symmetries are violated 
 not by the Higgs mechanism but by 
 the nontrivial boundary conditions. 
We have calculated the amplitudes of the process 
 of including four massive gauge bosons in the external lines. 
The results 
 show the power behavior of 
 both $O(E^4/m^2)$ and $O(E^2/m^2)$ 
 in the amplitude vanish. 
The calculations have be done 
 in the 5 dimensional 
 SM and the $SU(5)$ GUT, 
 whose 5th dimensional coordinate 
 is compactified 
 on $S^1/Z_2$. 
Therefore, 
 the broken gauge theory through 
 the orbifolding preserves unitarity 
 at high energy. 
This 
 situation is similar to the 
 broken gauge theory 
 where the gauge bosons obtain their 
 masses through the Higgs mechanism. 
This result shows that 
 the KK-modes in 
 the extra dimensional theory where 
 the gauge symmetry is violated 
 through the boundary condition 
 play an important role as 
 the Higgs scalars do. 
The structure of the interactions 
 among KK states
 are crucial for conserving the unitarity.

Finally, we show that 
 the 5th gauge field, 
 $A_5 (x^\mu, y)$ can be 
 {\it gauged away} and absorbed into the 
 longitudinal component of the 4D gauge 
 field $A_\nu (x^\mu, y)$, 
 both of which are given previously as Eqs.(\ref{3})$\sim$(\ref{2.6}).
This is realized 
 through the appropriate 
 gauge transformation 
 such as 
\begin{eqnarray}
&&A_\nu (x^\mu, y) \rightarrow U(x^\mu, y) A_\nu (x^\mu, y) 
U(x^\mu, y)^{-1}- iU(x^\mu, y) \partial_\nu U(x^\mu, y)^{-1}, \\
&&A_5 (x^\mu, y) \rightarrow U(x^\mu, y) A_5 (x^\mu, y) 
U(x^\mu, y)^{-1}- iU(x^\mu, y) \partial_5 U(x^\mu, y)^{-1}, 
\end{eqnarray}
where 
\begin{eqnarray}
U(x^\mu, y)={\mathcal P} \exp [i \int A_5(x^\mu, y)dy]. 
\end{eqnarray}
Here ${\mathcal P}$ is the path ordered product. 
It is noted that the above 
 transformation is compatible with the boundary 
 conditions on the 
 $S^1/Z_2$ orbifold as 
\begin{eqnarray}
&&U(x^\mu, y+2\pi R)= U(x^\mu, y), \\
&&U(x^\mu, -y)= PU(x^\mu, y)P^{-1},
\end{eqnarray}
due to 
the relation on the 5th gauge field as
\begin{eqnarray}
&&A_5(x^\mu, y+2\pi R)= A_5(x^\mu, y), \\
&&A_5(x^\mu, -y)= -PA_5(x^\mu, y)P^{-1}.
\end{eqnarray}
The 5th gauge field 
 is 
 the {\it would-be NG-like} field. 
A detailed study along this line 
 of thought
 will be presented elsewhere\cite{f}.


\vskip 1.5cm

\leftline{\bf Acknowledgments}

We would like to thank H.\ Murayama and Y.\ Nomura for helpful 
 comments. 
This work was supported in part by  Scientific Grants from 
 the Ministry of Education and Science, Grant No.\ 14039207, 
 Grant No.\ 14046208, \ Grant No.\ 14740164.

\vskip 1.5cm

\def\jnl#1#2#3#4{{#1}{\bf #2} (#4) #3}

\def\Zphys{{\em Z.\ Phys.} }
\def\jssc{{\em J.\ Solid State Chem.\ }}
\def\jpsJ{{\em J.\ Phys.\ Soc.\ Japan }}
\def\ptps{{\em Prog.\ Theoret.\ Phys.\ Suppl.\ }}
\def\PTP{{\em Prog.\ Theoret.\ Phys.\  }}

\def\JMP{{\em J. Math.\ Phys.} }
\def\NPB{{\em Nucl.\ Phys.} B}
\def\NP{{\em Nucl.\ Phys.} }
\def\PLB{{\em Phys.\ Lett.} B}
\def\PL{{\em Phys.\ Lett.} }
\def\PRL{\em Phys.\ Rev.\ Lett. }
\def\PRB{{\em Phys.\ Rev.} B}
\def\PRD{{\em Phys.\ Rev.} D}
\def\PRe{{\em Phys.\ Rep.} }
\def\AP{{\em Ann.\ Phys.\ (N.Y.)} }
\def\RMP{{\
em Rev.\ Mod.\ Phys.} }
\def\ZPC{{\em Z.\ Phys.} C}
\def\SCI{\em Science}
\def\CMP{\em Comm.\ Math.\ Phys. }
\def\MPLA{{\em Mod.\ Phys.\ Lett.} A}
\def\IJMPA{{\em Int.\ J.\ Mod.\ Phys.} A}
\def\IJMPB{{\em Int.\ J.\ Mod.\ Phys.} B}
\def\EPJC{{\em Eur.\ Phys.\ J.} C}
\def\PR{{\em Phys.\ Rev.} }
\def\JHEP{{\em JHEP} }
\def\cmp{{\em Com.\ Math.\ Phys.}}
\def\JPA{{\em J.\  Phys.} A}
\def\CQG{\em Class.\ Quant.\ Grav. }
\def\ATMP{{\em Adv.\ Theoret.\ Math.\ Phys.} }
\def\ibid{{\em ibid.} }

\leftline{\bf References}

\renewenvironment{thebibliography}[1]
         {\begin{list}{[$\,$\arabic{enumi}$\,$]}  
         {\usecounter{enumi}\setlength{\parsep}{0pt}
          \setlength{\itemsep}{0pt}  \renewcommand{\baselinestretch}{1.2}
          \settowidth
         {\labelwidth}{#1 ~ ~}\sloppy}}{\end{list}}

\end{document}